\documentclass{article}

\usepackage{arxiv}

\usepackage[utf8]{inputenc} 
\usepackage[T1]{fontenc}    
\usepackage{hyperref}       
\usepackage{url}            
\usepackage{booktabs}       
\usepackage{amsfonts}       
\usepackage{nicefrac}       
\usepackage{microtype}      
\usepackage{lipsum}
\usepackage{graphicx}
\graphicspath{ {./images/} }

\usepackage{caption}
\usepackage{subcaption}
\usepackage{algorithm}
\usepackage{algcompatible}
\usepackage{multirow}
\usepackage{textcomp, gensymb}
\usepackage{url}
\usepackage{makecell}

\title{Lisbon Hotspots: Wi-Fi access point dataset for time-bound location proofs}

\date{August 5\textsuperscript{th}, 2022}	

\author{ 
  Rui Claro, Samih Eisa, Miguel L. Pardal  \\
  INESC-ID, Instituto Superior T\'{e}cnico, Universidade de Lisboa\\
  Rua Alves Redol, 9 -- 1000-029 Lisbon, Portugal\\
  \texttt{\{rui.claro, miguel.pardal\}@tecnico.ulisboa.pt, samih.eisa@inesc-id.pt} \\
}

\begin{document}

\maketitle

\begin{abstract}
Wi-Fi hotspots are a valuable resource for people \textit{on the go}, especially tourists, as they provide a means to connect personal devices to the Internet.
This extra connectivity can be helpful in many situations, e.g., to enable map and chat applications to operate  \textit{outdoors} when cellular connectivity is unavailable or is expensive.
Retail stores and many public services have recognized that hotspots have potential to attract and retain customers, so many of them offer \textit{free and open Wi-Fi}.
In busy cities, with many locals and visitors, the number of hotspots is very significant.
Some of these hotspots are available for long periods of time, while others are short-lived.
When we have many users with devices collecting \textit{hotspot observations}, they can be used to \textit{detect the location} -- using the long-lived hotspots -- and to \textit{prove the time when the location was visited} -- using the short-lived hotspots observed by others users at the location.

In this article, we present a \textit{dataset} of collected Wi-Fi data from the most important tourist locations in the city of Lisbon, Portugal, over a period of months, that was used to show the feasibility of using hotspot data for location detection and proof.
The obtained data and algorithms were assessed for a specific use case: smart tourism.
We also present the \textit{data model} used to store the observations and the \textit{algorithms} developed to detect and prove location of a user device at a specific time.

The \emph{Lisbon Hotspots} dataset, \emph{LXspots}, is made publicly available to the scientific community so that other researchers can also make use of it to develop new and innovative mobile and Internet of Things applications.
\end{abstract}


\section{Introduction and Related Work}
\label{sec:introduction}

Discovering an individual's location \textit{on the go} is a common and indispensable function of the smartphones and many wearable devices.
Location-based services use geographical information to provide useful applications for the end-users such as: maps, driving assistance, parking, goods delivery, and trip advisory.
For outdoor location, the dominant solution is to use GPS \cite{zeng2017practical, Hameed2018SurveyOI}.
For indoor location, there are solutions that use signal strength collected from Wi-Fi Access Points (APs) to determine the locations inside specific buildings~\cite{haeberlen2004practical}, after a complete mapping of the signal strength in each room has been created earlier.
However, in both outdoors and indoors, the user's location is vulnerable to hacking and spoofing attacks~\cite{lee2010location}, leaving some services unprotected from malicious users who fake their locations~\cite{zeng2017practical,humphreys2012statement,tippenhauer2012iphone}.
Location proofs provide protection against these types of attacks by creating digital certificates that attest to an individual's presence at a geographical location whereby services can validate the location claim.

\subsection{Location Proofs}

Location proof (or location certification) systems collect evidence when a device is at a specific location.  The evidence can be stored and then later be verified, proving the device was at a specific time and place.
For example, the STAMP~\cite{wang2016stamp} system provides time-bound location proofs where mobile users can generate proofs for each other.
%
There are also solutions for moving vehicles, such as the Vouch system~\cite{boeira2018vouch}.
The SureThing system~\cite{ferreira2018witness} also allows devices to produce and validate location proof certificates, to make proof of their locations and to reliably verify the locations of other devices, using the neighboring devices as witnesses that as well collect GPS, Wi-Fi, or Bluetooth evidence.

More recently, SureThing was expanded to become a framework~\footnote{\url{http://surething-project.eu}}.
It now provides common data formats and procedures to be used by applications that use location proofs.
It allows system participants to play different roles, with flexibility.
The \textit{prover} role is usually played by a device that makes a location claim backed by some evidence.
A \textit{witness} is another device that endorses the claim and adds its own evidence.
Finally, a \textit{verifier} device (usually a server) analyzes all the evidence and ultimately makes a decision to issue - or not - a location certificate.
Each application has its own operator that assigns the roles and the authority of the verifiers.
This allows each application to decide which specific kinds of location evidence need to be collected and presented by the provers and witnesses, and also what are the specific conditions for the verifiers to accept a location claim and issue a corresponding location certificate.

The SureThing framework supports both ad-hoc and fixed witnesses.
On the one hand, \textit{ad-hoc witnesses} are neighboring devices that verify the location of the prover.
These witnesses need not to be fully trusted, since they are not directly controlled by the system operator (or by the verifier) and their security comes more from their quantity and diversity, \textit{i.e.}, there need to be multiple and different witnesses to support a location claim.
On the other hand, \textit{fixed witnesses} are devices that are placed on-site by the operator and can be more trusted.
These can take the form of kiosks or other dedicated hardware.

The SureThing framework also supports \textit{beacons}.
These devices can be added on-site and, once deployed, broadcast unique signals that can be picked up by provers and witnesses.
The signals can be random, or pseudo-random sequences.
In this latter case, the verifier can predict the signal values, if it knows the seed number, and the overall system is kept in synchrony, \textit{i.e.}, clock skew below a maximum difference time for all devices \cite{Tiago22}.

Ideally, to keep its costs down and availability up, a location proof application should rely as much as possible on existing devices.
If these devices are owned and operated by third parties, then we can have a signal \textit{scavenging} approach to build a location proof.

\subsection{Wi-Fi Scavenging}

A scavenging strategy for location proofs is centered around the idea of collecting existing signals at public places like retail stores, restaurants, and public services.
A device does not need to connect to the network, as it only needs to see the announced identifiers, like the SSID (service set identifier).
If viable, this strategy requires minimal investment, since only previously available infrastructure is used, whose operation cost is already being supported, usually by third parties.

Most cities nowadays have plenty of Wi-Fi hotspots available for public use.
The scavenging approach is promising because the hotspot networks may be divided in two sets: a set of networks that remain available over long periods of time and another set that change more frequently.
The former are likely associated with retail stores and services whereas the latter are probably associated with vehicles and people passing by the location.
The idea then, is to take advantage of the long-lived hotspots to \textit{detect the location} and to use the short-lived hotspots to \textit{prove the time when the location was visited}. 
Wi-Fi traces can be captured by the user device at the visited locations and compared later with traces collected by devices of other users that were co-located at the same locations.
This approach is only expected to work in busy locations, so that the short-lived hotspots are sufficient in number for the desired time span of the location proof.

\subsection{Field work}

To validate the hypothesis of using Wi-Fi signals collected from public hotspots for location proofs, we set out to do field work for collecting data and then verifying if the approach was feasible.
We chose the city of Lisbon, Portugal for the real-world data collection.
The work occurred over a period of 6 months, on 6 locations, and we made a total of 11 data collection sessions in all locations.
We picked locations of interest, namely tourist attractions, because of their large number of wireless networks and potential high number of available ad-hoc witnesses at any given period of time.

A prover device creates evidence for a location by collecting Wi-Fi signals from nearby APs at a single location.
The APs are identified by their associated unique SSID (Service Set IDentifier) and other Wi-Fi signal characteristics.
On the verifier, the evidence from the location claim is  compared against previously stored location evidence, submitted by other devices that act as witnesses.
For a time-bound location proof, the verifier tries to establish a time interval with evidence of co-location of the prover device and its witnesses.
This gives the verifier the ability to ensure that the prover location claim is valid within certain time-interval.

\subsection{Contributions}

The contributions of this work are the following:
\begin{itemize}
    \item A Wi-Fi access point dataset collected in the city of Lisbon, Portugal;

    \item A data model 
    to store and query the collected observations;

    \item Algorithms to determine the location and time bounds of the visits that can be used to issue location certificates.
\end{itemize}

\subsection{Document Overview}

The rest of the paper is organized as follows:
Section~\ref{sec:dataset} presents the data set and how it was collected;
Section~\ref{sec:assessment} analyses the dataset in the context of a \textit{smart tourism} use case;
Section~\ref{sec:model} presents a formal model of data and algorithms, defined to make inferences from the dataset.
The paper concludes in Section~\ref{sec:conclusion}.

\section{Dataset}
\label{sec:dataset}

We performed a field experiment to collect Wi-Fi access point traces.
The goal is to use this data to later assess the viability of the scavenging approach to produce location proofs.
The dataset is called \textit{Lisbon hotspots} or just \textit{LXspots} and it is publicly available \footnotemark.
We present the rationale for selecting locations, the collection sessions, and the details of the collected data.
\footnotetext{\url{https://github.com/inesc-id/SureThing-LXspots}} 

\subsection{Location Selection}

We started by selecting the locations where data was going to be collected.
The locations to select should contain different types of attractions, while also containing different types of Wi-Fi networks.
Our selection was based on the following criteria: \emph{Indoor vs Outdoor}, \emph{Dense vs Sparse} and \emph{Central vs Remote}.
Indoor locations tend to have more variation in Wi-Fi signal strength when comparing to Outdoor locations, since more sources of interference exist.
Also indoor locations are more likely to have higher number of Wi-Fi APs than outdoor locations.
The population density \footnotemark on a location is reflected based on the types of captured Wi-Fi networks.
\footnotetext{In this work, population density describes the volume of people visiting/passing near a collection location, and it is not related to the statistical index of population per unit area.} 
Highly populated areas tend to have more Wi-Fi mobile hotspots.
On the contrary, sparsely populated ones tend to have more fixed Wi-Fi APs.
Finally, the actual position of attractions in the city influences the collected Wi-Fi traces from the APs.
The locations that are more central in the city tend to have more Wi-Fi APs and more likely to have higher population density than the remote locations.

Once the criteria were set, we used well-known traveling websites to retrieve the top tourist attractions places recommended for the tourists visiting the city of Lisbon.
Namely, we used: TripAdvisor, Booking, and City Tour bus lines.
We then filtered the locations from those websites to get only 5 that better fulfilled the criteria identified above.
Finally, we added one extra location that represents a residential area (Reference Name: \textit{Alvalade}), so that we could observe differences between the attractions and residential neighbourhoods.

\begin{figure}[htp]
\centering
\begin{tabular}{cc}
  \includegraphics[width=37mm, height=37mm, angle =0]{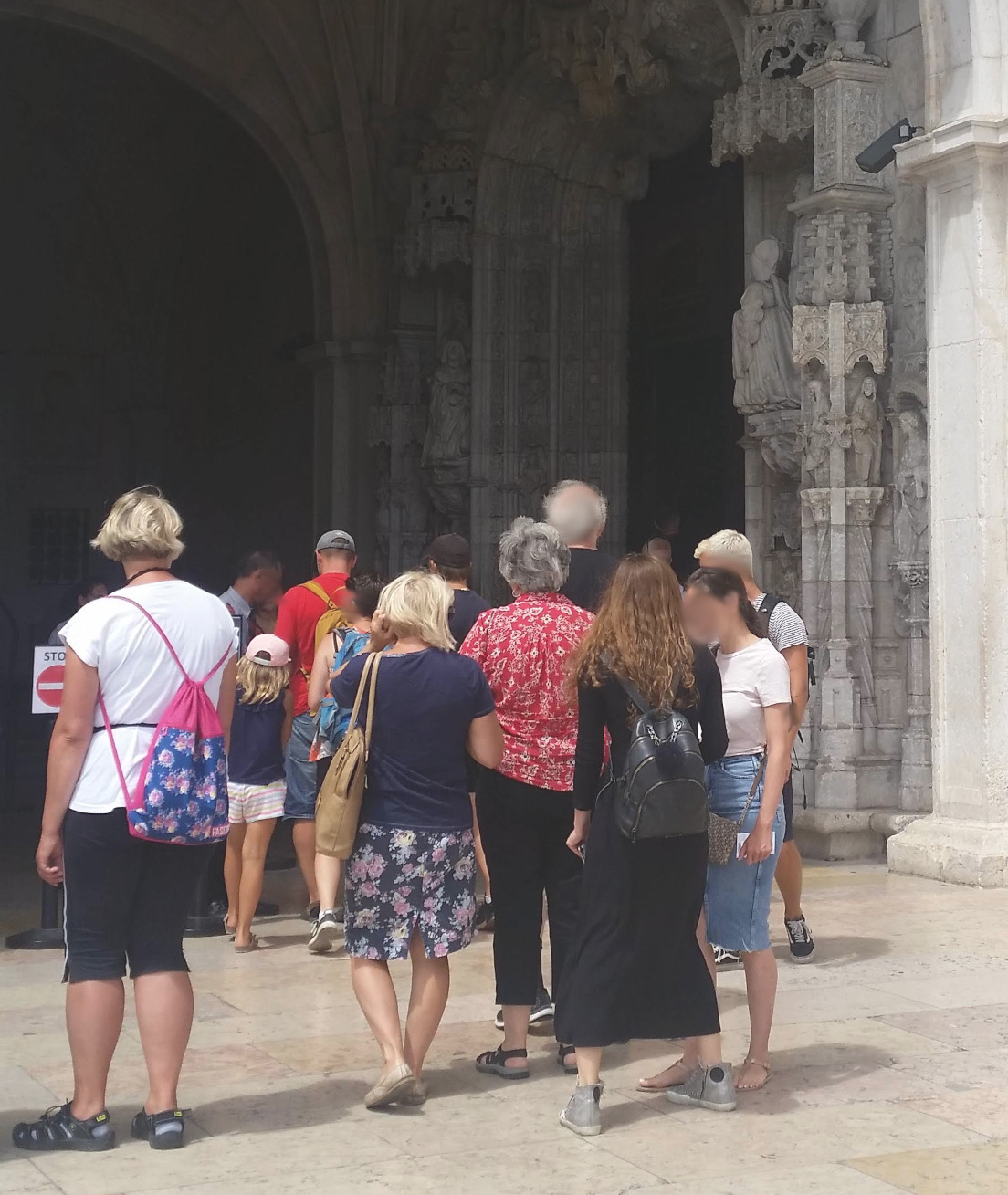} &   \includegraphics[width=37mm, height=37mm, angle =0]{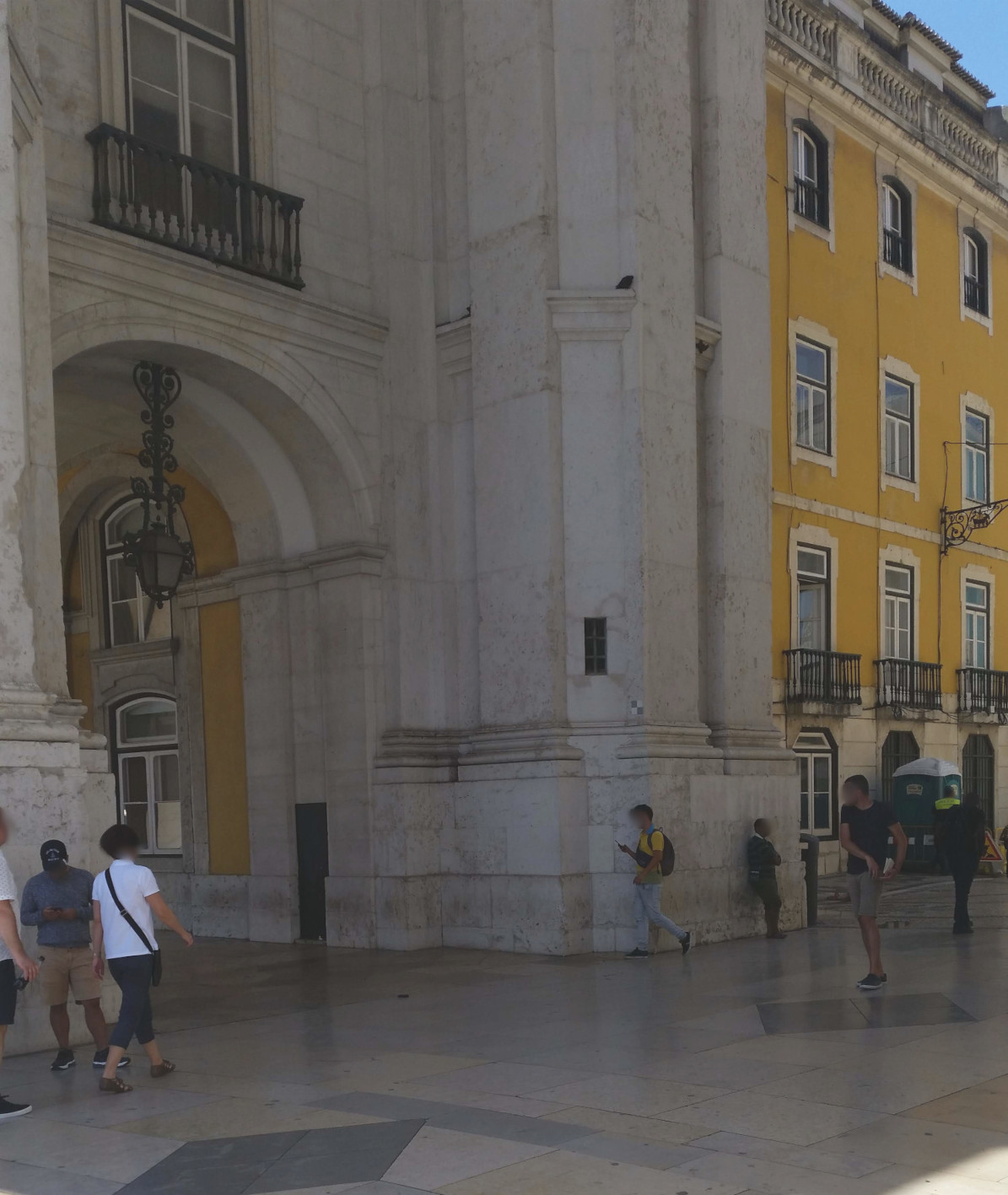} \\
(a) Jer\'{o}nimos & (b) Com\'{e}rcio \\[6pt]
 \includegraphics[width=37mm,height=35mm, angle =0]{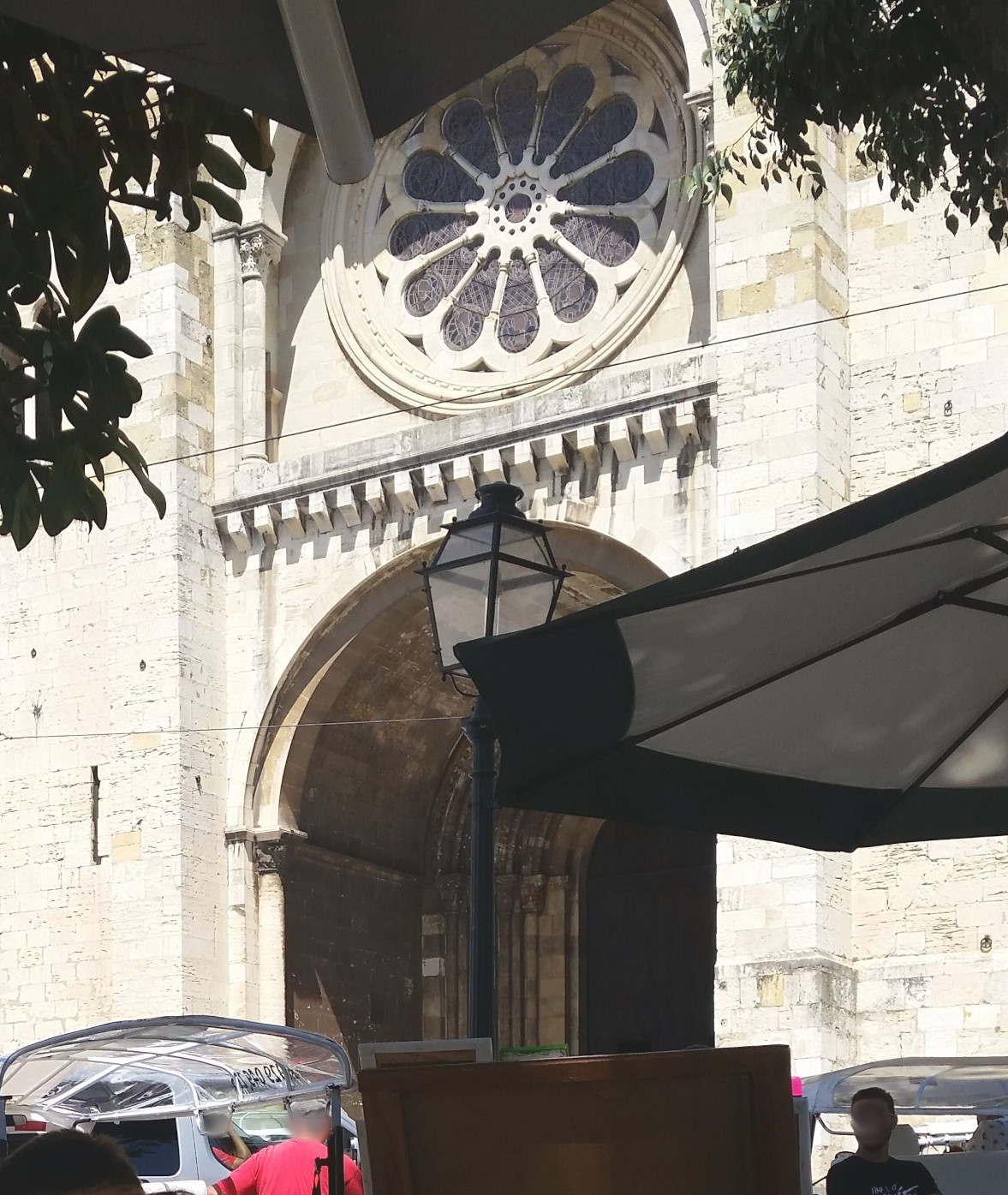} &   \includegraphics[width=37mm, height=37mm, angle =0]{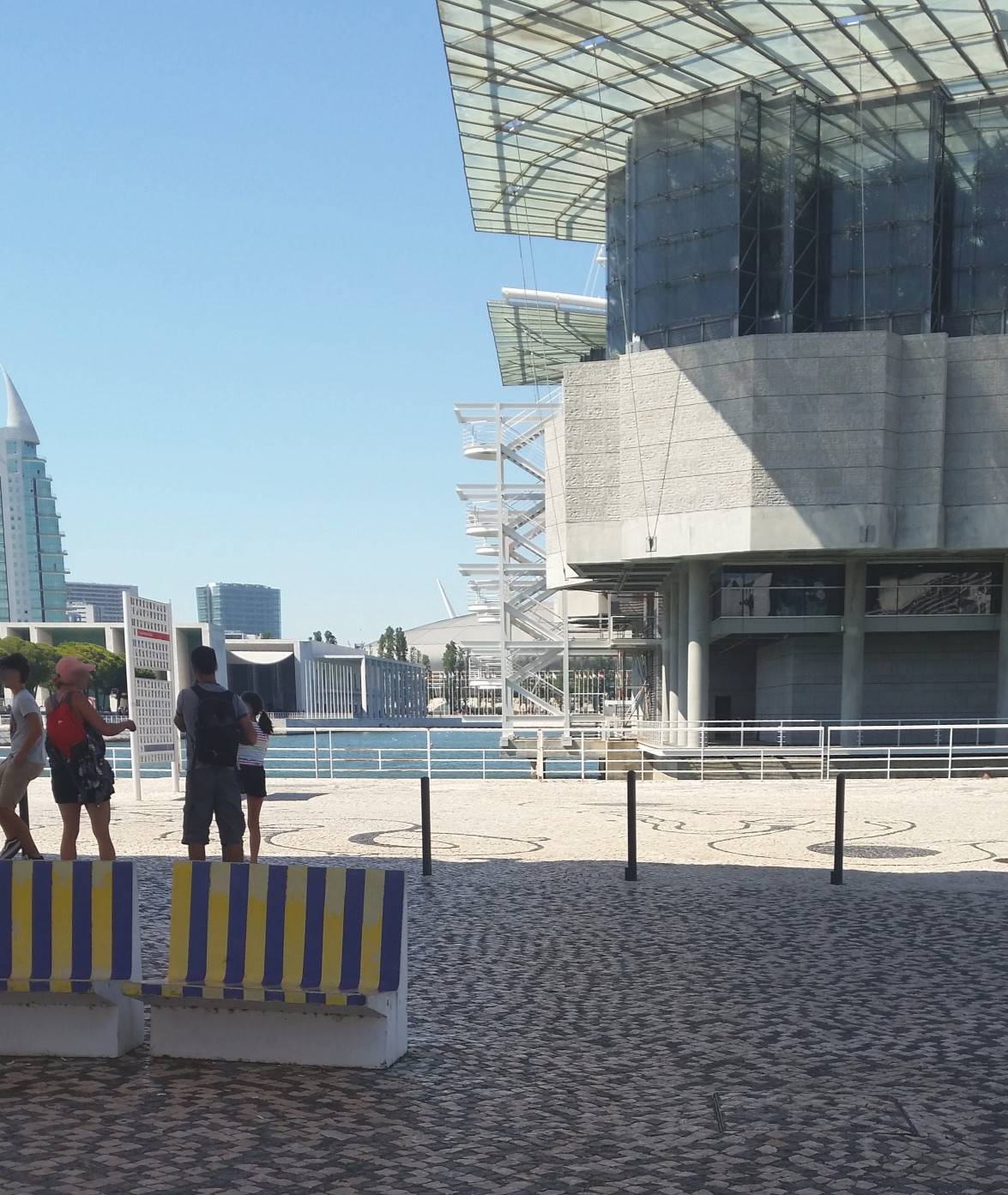} \\
(c) S\'{e} & (d) Ocean\'{a}rio \\[6pt]
\includegraphics[width=37mm, height=37mm, angle =0]{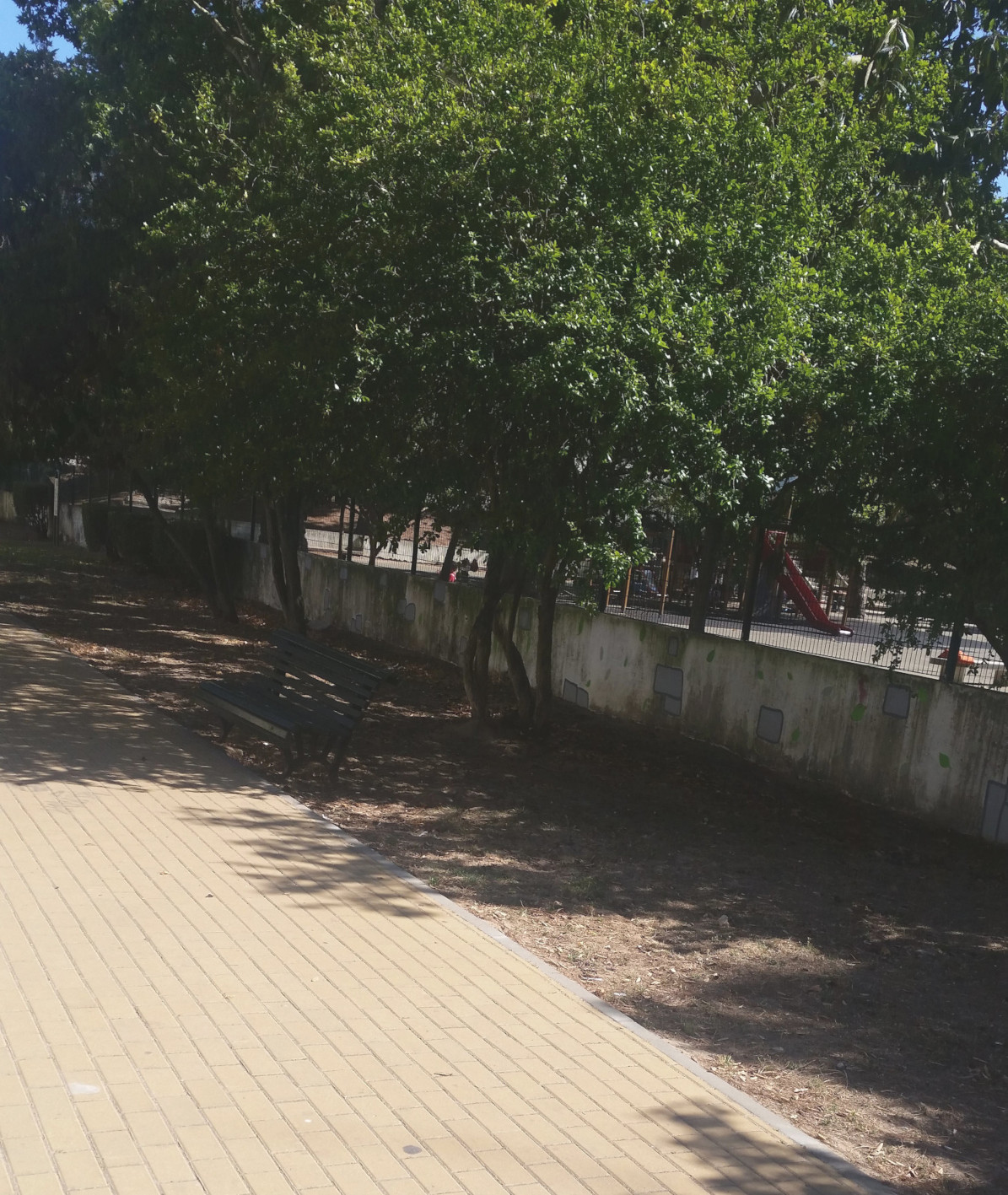} &   \includegraphics[width=37mm, height=37mm, angle =0]{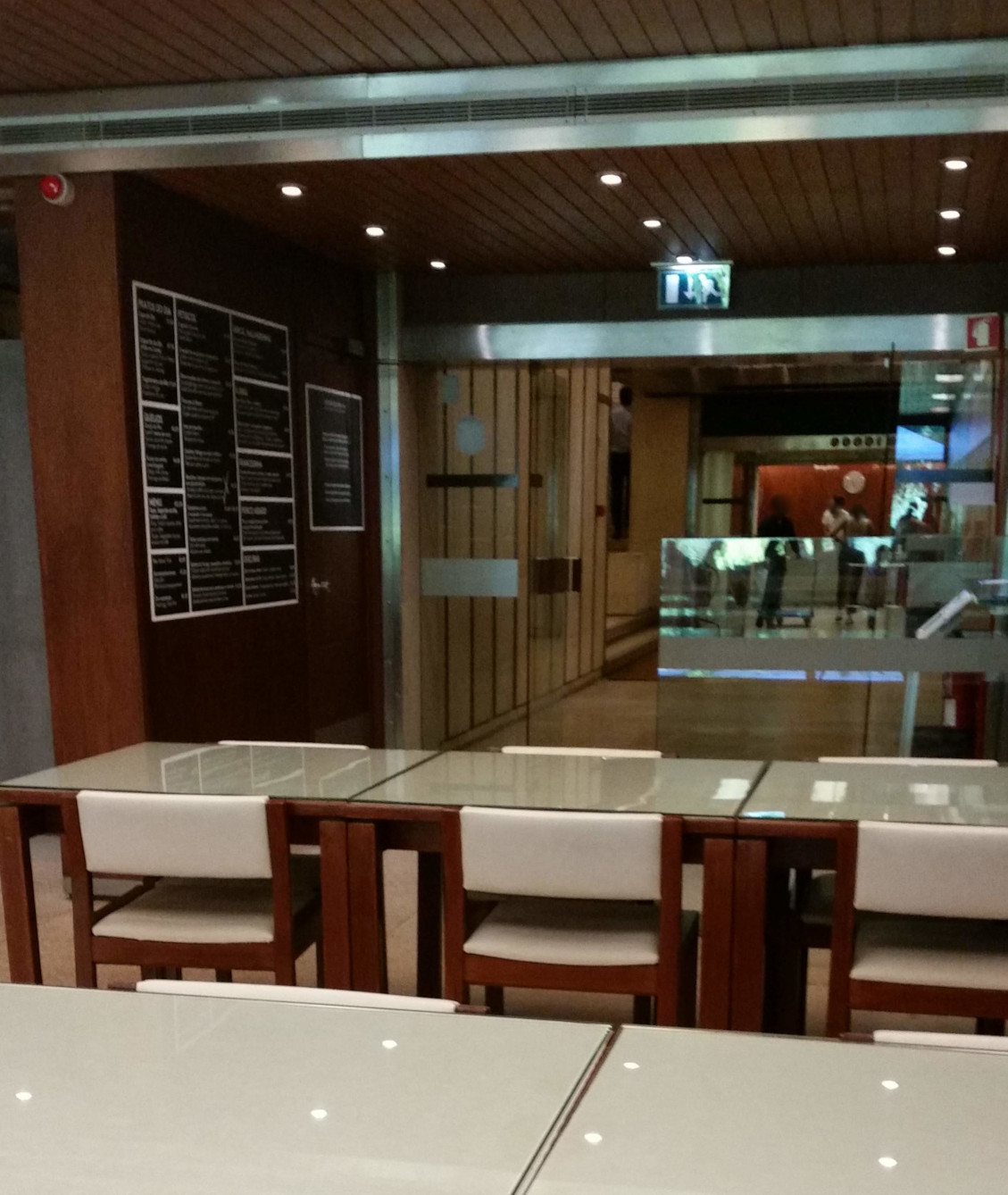} \\
(e) Alvalade & (f) Gulbenkian \\[6pt]
\end{tabular}
\caption{Pictures taken at each of the LXspots data collection locations.}
\label{fig:LX_pictures}
\end{figure}

The selected locations are listed in Table~\ref{table:locations_LX} and 
Figure~\ref{fig:LX_pictures} shows a photo taken at each of the locations.


\begin{table}[htp]
	\caption{Data collection locations of the LXspots dataset.}
	\label{table:locations_LX}
	\centering
	\resizebox{0.60\columnwidth}{!}{%
	\begin{tabular}{|c|l|l|l|}
		\hline
		& Reference Name     & Coordinates                              & Matched Criteria   \\ \hline
		A & Jer\'{o}nimos      & $38\degree41'50.9''N 9\degree12'21.5''W$ & Outdoor \& Dense   \\ \hline
		B & Com\'{e}rcio       & $38\degree42'30.2''N 9\degree08'12.1''W$ & Central \& Dense   \\ \hline
		C & S\'{e}             & $38\degree42'34.8''N 9\degree08'01.3''W$ & Central \& Outdoor \\ \hline
		D & Ocean\'{a}rio      & $38\degree45'45.1''N 9\degree05'38.7''W$ & Remote  \& Outdoor \\ \hline
		E & Alvalade           & $38\degree45'16.4''N 9\degree08'48.3''W$ & Remote  \& Sparse  \\ \hline
		F & Gulbenkian         & $38\degree44'13.7''N 9\degree09'12.0''W$ & Central \& Indoor  \\ \hline
	\end{tabular}}
\end{table}

A total of six (6) different locations were selected across the city of Lisbon, Portugal.
Five (5) of them reflect highly visited tourism attractions such as museums and cathedrals.
The 6th location was intentionally a residential area of the city, to see how the Wi-Fi networks are different in a non-touristic location.

\subsection{Data Collection}

Data was collected at each location over 6-month period, with most of the collection concentrated during 1 week.
The collected data is composed of discrete measurements of existing Wi-Fi networks.
The measurements contain detailed information obtained through Wi-Fi scanning such as MAC addresses and signal intensities.

As mentioned, the data collection was done over 6-month period, and since we targeted public places, continuous scavenging was not possible due to legal and infrastructural constraints.
Our approach was to visit each location, during the course of a day, and gather data for a time span of 15 minutes.
The visit route was settled from location \textit{A} to \textit{F} for ease of navigation through the city.
The first collection route was on July 19th 2019, and the last was on January 19th 2020.
Table~\ref{table:days_LX} details each of the days and the rationale for selecting them.

\begin{table}[h]
	\caption{Days of data collection for the LXspots dataset.}
	\centering
	\label{table:days_LX}
	\setlength{\tabcolsep}{3pt}
	\begin{tabular}{|l|l|l|}
		\hline
		ID   & Day                      & \multicolumn{1}{c|}{Observations}  \\ \hline
		1    & 2019-07-19               & First day of scavenging.           \\ \hline
		2    & 2019-07-26               & One week after first scavenging.   \\ \hline
		3..9 & 2019-07-29 : 2019-08-04  & Full week of scavenging.           \\ \hline
		10   & 2019-08-19               & One month after first scavenging.  \\ \hline
		11   & 2020-01-19               & Six months after first scavenging. \\ \hline
	\end{tabular}
\end{table}

For redundancy, the data collection was done using three different smartphones.
Each one has a scavenger mobile application installed to detect nearby Wi-Fi networks and retrieve their properties.
The application was installed on three different smartphones running the Android operating system: Samsung Galaxy S9, Huawei Mate 10, and LG V10 thinq.
We will refer to these smartphones in the rest of the paper as devices A, B and C, respectively.

\subsection{Data Features}
\label{subsection:dataset}

The majority of the data features describe information related to the Wi-Fi network protocol.
Additionally, there are features that present information related to the GPS position, date and time of collection, the device used, and reference names of the locations.
The full details of the features are presented in Table~\ref{table:features_LX}.

\begin{table}
	\caption{Features present on the LXspots dataset.}
	\label{table:features_LX}
	\centering
	\resizebox{0.875\columnwidth}{!}{%
	\begin{tabular}{|l|l|}
		\hline
		\multicolumn{1}{|c|}{Feature Name} & \multicolumn{1}{c|}{Description}                                                                                                                                                                                                            \\ \hline
		device\_id                         & Device identifier {[}A,B or C{]}.                                                                                                                                                                                                           \\ \hline
		date                               & Date of the observation.                                                                                                                                                                                                                    \\ \hline
		time                               & Time of the observation.                                                                                                                                                                                                                    \\ \hline
		ref\_name                          & Location reference name.                                                                                                                                                                                                                    \\ \hline
		latitude                           & Latitude in degrees.                                                                                                                                                                                                                        \\ \hline
		longitude                          & Longitude in degrees.                                                                                                                                                                                                                       \\ \hline
		altitude                           & Altitude in meters above the WGS 84 reference ellipsoid.                                                                                                                                                                                    \\ \hline
		accuracy                           & Estimated horizontal accuracy, radial, in meters.                                                                                                                                                                                           \\ \hline
		SSID                               & Service Set IDentifier, the network name.                                                                                                                                                                                                                           \\ \hline
		BSSID                              & Basic Service Set IDentifier, the address of the access point.                                                                                                                                                                                                            \\ \hline
		capabilities                       & Authentication, key management, and encryption schemes supported.                                                                                                                                                       \\ \hline
		frequency                          & The primary frequency of the channel {[}MHz{]}.                                                                                                                                                                                             \\ \hline
		level                              & The detected signal level in dBm, also known as the RSSI (Received Signal Strength Indicator).                                                                                                                                                                                   \\ \hline
		centerfreq0                        & \begin{tabular}[c]{@{}l@{}}0 if AP bandwidth is 20 MHz.\\ If the AP uses 40, 80 or 160 MHz, center frequency {[}MHz{]}.\\ AP use 80 + 80 MHz, center frequency of the first segment {[}MHz{]}.\end{tabular} \\ \hline
		centerfreq1                        & AP use 80 + 80 MHz, center frequency of the second segment {[}MHz{]}.                                                                                                                                                   \\ \hline
		channelwidth                       & Channel bandwidth {[}0=20MHz; 1=40MHz; 2=80MHz; 4=160MHz{]}.                                                                                                                                                                       \\ \hline
	\end{tabular}}
\end{table}

\subsection{Summary}

For the LXspots dataset, a total of 6 different locations across the city of Lisbon, Portugal, were selected;
5 of them reflect highly visited tourism attractions such as museums and cathedrals.
The data was collected using multiple mobile devices and over different days of the year, during a busy tourism season and in an almost standstill of a city-wide lockdown\footnotemark.
\footnotetext{Specifically, data collection was done during the months of July 2019, January 2020 and July 2020; with the last one done during the city lockdown caused by the COVID-19 pandemic.}
The most important data features are the GPS position, the date and time of collection, the device used, and the reference identifiers of the locations.

\section{Assessment}
\label{sec:assessment}

In this section we leverage the collected dataset and assess the feasibility of using a scavenging approach to location proofs with time-bounds for a specific use case: smart tourism.

\subsection{Smart Tourism Use Case}
\label{sec:usecase}

We will assume a smart tourism application~\cite{maia2020cross} as background for the feasibility assessment.
Smart tourism is an important byproduct of a smart city ecosystem.
This new approach to traditional tourism has greatly benefited from technological innovation, with new applications appearing in different business fields~\cite{gretzel2015smart}.
More specifically, we think that the main benefit is routing people from main tourist attractions to less-known ones, promoting better distribution of visitors to decongest popular attractions.


\begin{figure}
	\centering
	\includegraphics[width=0.8\columnwidth]{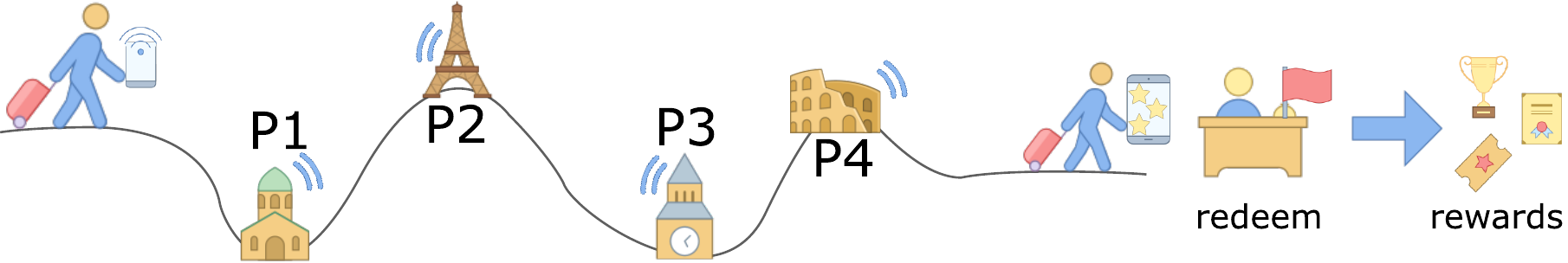}
	\caption{The smart tourism application rewards users that collect location proofs in all locations~\cite{Maia_2019_INForum_CROSS}.}
	\label{fig:tourism_route}
\end{figure}

We assume that each tourist will carry its mobile phone running the application that is collecting the Wi-Fi hotspots.
The application offers a small reward, like a souvenir or a discount coupon, to each user that visits all the locations in a tourism route, as 
illustrated in Figure~\ref{fig:tourism_route}.

\begin{figure}
	\centering
	\includegraphics[width=0.70\columnwidth]{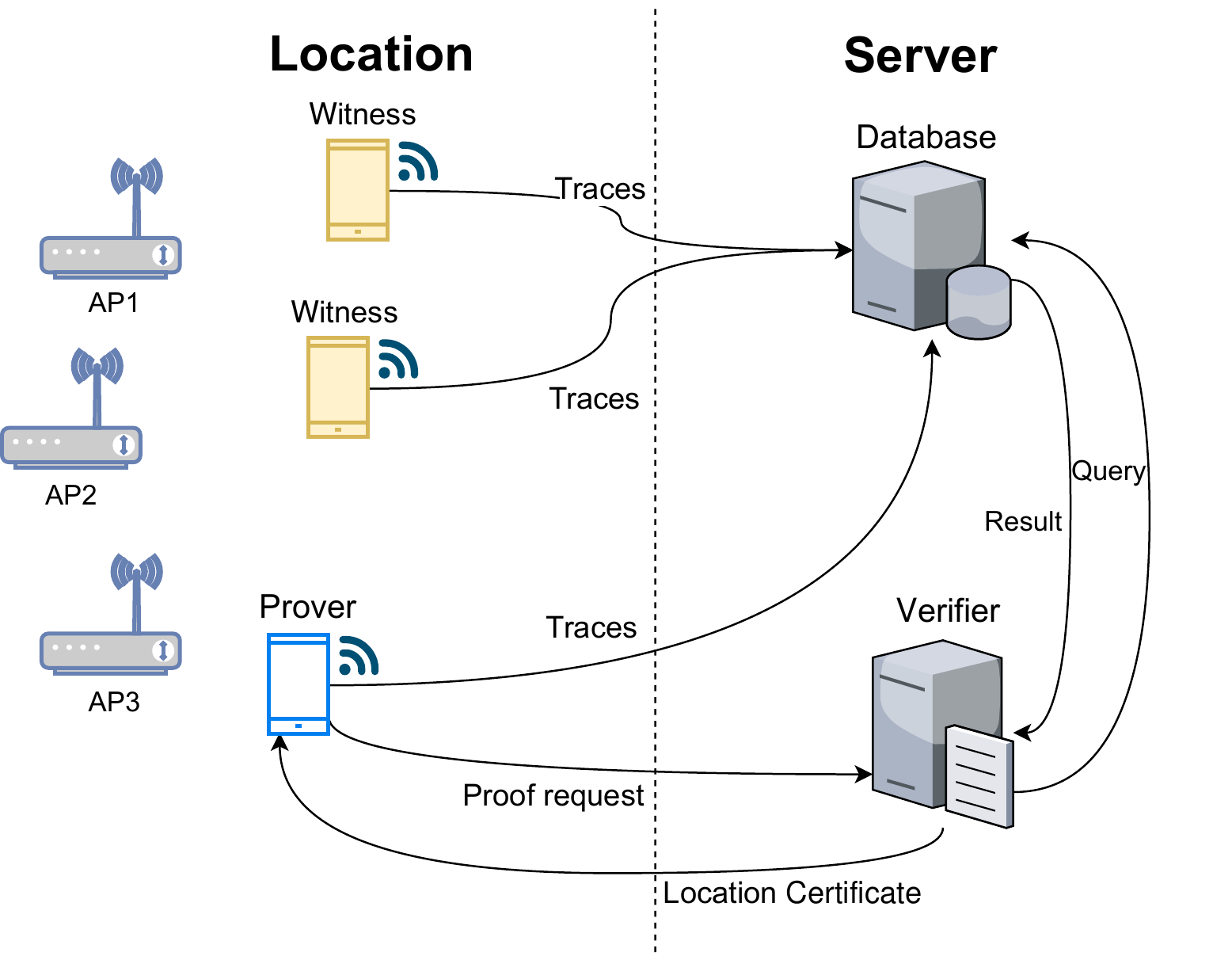}
	\caption{Location proof system based on scavenged Wi-Fi access point data.}
	\label{fig:scavenging_timebound_proof}
\end{figure}

The overall assumed system is represented in Figure~\ref{fig:scavenging_timebound_proof}.
The APs are broadcasting their identifiers.
The mobile devices are collecting the Wi-Fi traces and uploading them to the application server.
The prover device also uploads its traces and, when it needs a location certificate, it sends a request with the claimed location and time to the verifier.
The verifier accesses the database and checks if there is enough evidence to certify the prover location in the claimed time (or interval) .
If so, a location certificate is issued and returned to the prover.

\subsection{Data Processing}

We will process the data collected in multiple sessions to compute the long-lived hotspots -- that we call \emph{stable networks} -- and the short-lived ones -- that we call the \emph{volatile networks}.
The collected data was divided into training and testing sets.
The first dataset was comprised of the first 10 collection days.
The testing dataset contains data collected 6 months after the initial one.

\subsubsection{Stable Networks}
\label{subsec:eval_stable}

The training dataset is used to identify the stable Wi-Fi networks at the locations.
This dataset contains in total 10 days of data collection.

The first step was to merge the observations at each location from all the 10 days, from all the three devices (A, B, C).
This allowed us to count the total number of occurrences of each network in each place.
We then selected the top 10\% networks based on this count for each location.
This threshold was arbitrarily chosen in an ad-hoc way, given the values present in the dataset.

Table~\ref{table:stable_set} lists the total number of Wi-Fi networks present in each location and the number of calculated stable networks.
As expected, locations that were identified as densely populated (\textit{Jer\'onimos} and \textit{Com\'ercio}) have higher variety of networks when compared to residential area (\textit{Alvalade}).

\begin{table}
	\caption{Detected networks and stable set size for each location.}
	\label{table:stable_set}
	\centering
	\begin{tabular}{|l|c|c|}
		\hline
		\textbf{Location} & \textbf{Stable set} & \textbf{Total} \\ \hline
		Jer\'onimos         & 70                  & 677            \\ \hline
		Com\'ercio          & 58                  & 551            \\ \hline
		S\'e                & 47                  & 363            \\ \hline
		Ocean\'ario         & 25                  & 243            \\ \hline
		Alvalade          & 17                  & 163            \\ \hline
		Gulbenkian        & 30                  & 292            \\ \hline
	\end{tabular}
\end{table}

\begin{figure}
	\centering
	\includegraphics[width=0.80\columnwidth]{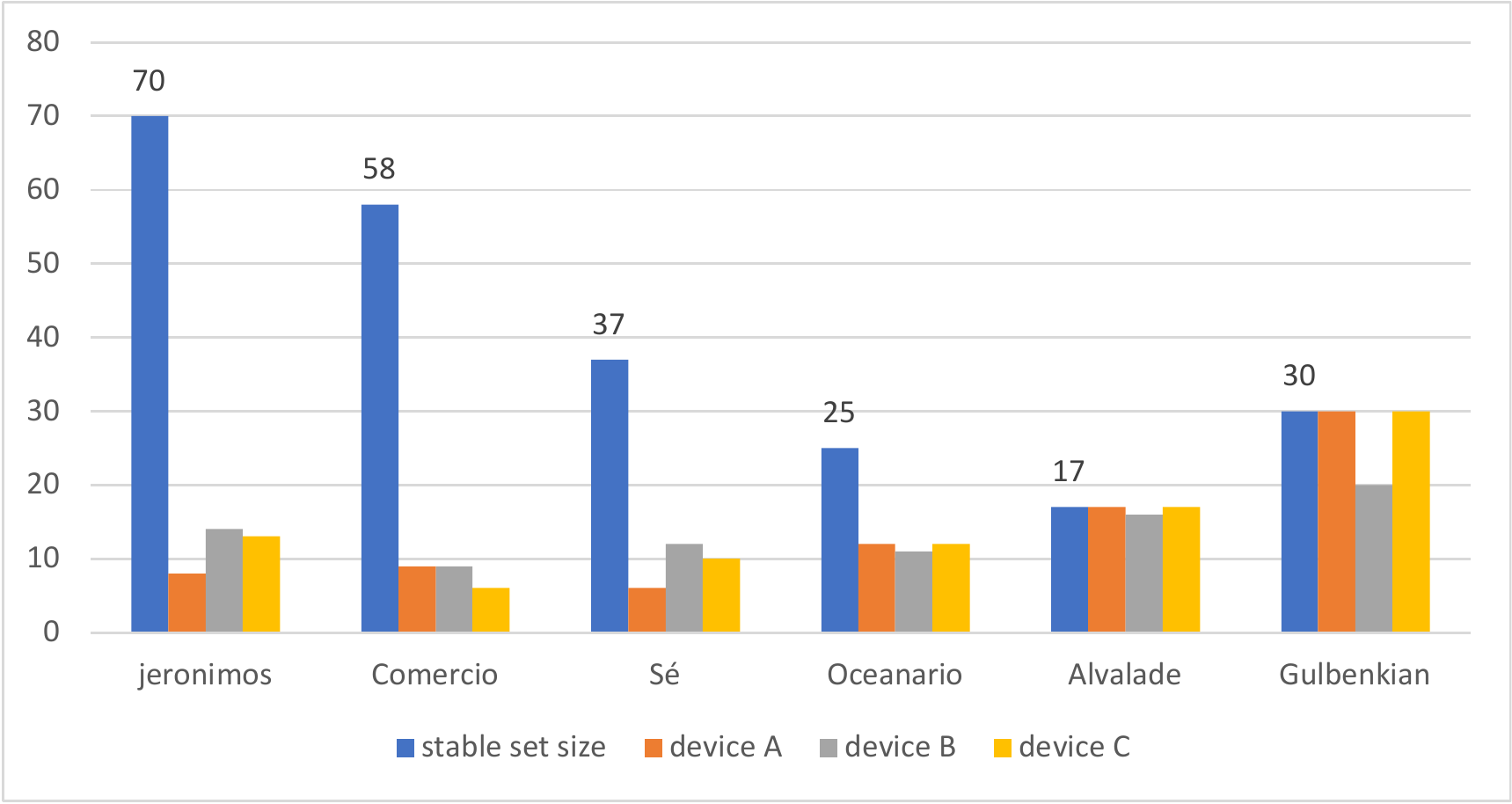}
	\caption{Stable set identification, 6 months after.}
	\label{fig:stable_set_chart}
\end{figure}

The second step was to verify that the calculated number of stable Wi-Fi networks could be detected by the prover's device.
For that, we used the testing dataset (as described before).
We separated the observations by each device and compared that with the stable networks, computed in the training step.
Figure~\ref{fig:stable_set_chart} presents the results.
The results show that we were able to identify, in all the six (6) locations, the networks that are present in the stable Wi-Fi networks set, with some disparities in the number of detected APs.
We identified the possible reason for these disparities: The type of locations has an impact in the immutability of the scavenged Wi-Fi networks.
For example, we have better results in \textit{Alvalade} (98\% matched) and in \textit{Gulbenkian} (89\% matched) than in \textit{Jer\'onimos} (14\% matched).
\textit{Alvalade} is a residential neighbourhood, and so has a large number of domestic APs owned by families.
These tend to remain stable through large period of time.
\textit{Gulbenkian} is an interesting location.
We identified that the networks contained in the stable networks set are almost only alias of three networks owned by the museum.
Moreover, since the data collection was done indoors, this was expected because we captured the Wi-Fi signals from multiple APs with the same SSID.
These institutions owned networks that also tend to remain stable.
On the other hand, \textit{Jer\'onimos} is an outdoor place, without many buildings nearby, reducing the number of stationary APs.
Thus, despite the larger number of networks detected, most of them were Wi-Fi hotspots.

\subsubsection{Volatile Networks} 
\label{subsec:eval_volatile}

The main purpose is to produce time-bound location proofs.
We now present the methodology to identify the volatile Wi-Fi networks set and the analysis that was done to validate the creation of time-bound location proofs.
In our approach, the volatile networks set is comprised of the bottom 10\% networks that were observed by a single device during a period of time.
Again, the specific threshold was chosen ad-hoc.
These networks, counted in the same manner as before, represent the total number of occurrences of each network.
We selected the bottom ones since those are the least observed, allowing us to shorten the time-period of the proofs.
We also removed from the volatile networks set the networks that are present in the stable networks set.
This was done as precaution to not falsely the produced location proof using the stable networks.

To create a time-bound location proof, both the prover and witnesses have to be in the same place, at the same time.
Using the 3 devices (A, B, C) in the dataset, we combined them in pairs, where one has the role of the prover, and the other of the witness.
We then generated the volatile networks set for each of the devices.
We compared the generated volatile networks set of the prover with the one generated by the witness.
This procedure was repeated for each pair of prover/witness, for each location in the dataset, and for 4 different time intervals (\emph{deltas}).
Table~\ref{table:volatile_set_devices} presents the results of the volatile networks identification.

\begin{table}
	\caption{Volatile networks detection for each location. Values indicate the number of pairs (prover/witness) that succeeded (out of 6 total).}
	\label{table:volatile_set}
	\centering
	\begin{tabular}{|l|r|r|r|r|}
		\hline
		\textbf{Location} & \textbf{15 min} & \textbf{7.5 min} & \textbf{3.75 min} & \textbf{1.875 min} \\ \hline
		Jer\'onimos       & 6               & 6                & 5                 & 3                  \\ \hline
		Com\'ercio        & 5               & 1                & 0                 & 0                  \\ \hline
		S\'e              & 6               & 4                & 2                 & 0                  \\ \hline
		Ocean\'ario       & 6               & 5                & 0                 & 0                  \\ \hline
		Alvalade          & 6               & 6                & 6                 & 4                  \\ \hline
		Gulbenkian        & 6               & 6                & 6                 & 6                  \\ \hline
	\end{tabular}
\end{table}

We considered a match if at least one volatile network is present on both the prover and witness sets.
We divided the 15 minute samples for each location and device to study how fine grained the temporal resolution can be.
The values present on the table refer to the number of prover/witness pairs (out of 6 total) succeeded in detecting at least one network on each other's volatile networks set.
We can see that for the 15-minute interval, almost all the pairs produce a match, with match percentage of 97\%.
As expected, these values decrease as we shorten the time interval, with the following match percentages: 97\% for the full 15-minute interval; 78\% for the 7.5-minute intervals; 53\% for the 3.75-minute interval; and 36\% for the 1.875-minute interval.
We also verified if the variation in values depended on the device that had the role of the prover.
However, as shown in Table~\ref{table:volatile_set_devices} the results are not dependant on the device.

\begin{table}
	\caption{Volatile networks detection for each device. Values indicate the number of pairs (prover/witness) that succeeded (out of 12 total.)}
	\label{table:volatile_set_devices}
	\centering
	\begin{tabular}{|l|l|l|l|l|}
		\hline
		\textbf{Device} & \textbf{15 min} & \textbf{7.5 min} & \textbf{3.75 min} & \textbf{1.875 min} \\ \hline
		A               & 11              & 10               & 6                 & 4                  \\ \hline
		B               & 12              & 10               & 7                 & 5                  \\ \hline
		C               & 12              & 8                & 6                 & 4                  \\ \hline
	\end{tabular}
\end{table}

\subsection{Discussion}

The results show that the stable networks set detection is sufficient for the smart tourism use case, but if we want to add more guarantees to the location proof, stronger constrains need to be placed.
Instead of detecting only a percentage of stable networks, an alternative would be to detect all networks present in the stable networks set.
This alternative gives stronger guarantees, but raises new challenges, for example, requiring more intervention from the system operator if an AP from the stable networks is physically removed.

From our experimental analysis with the volatile networks, we observe that for intervals of approximately 7 minutes, our approach can produce time-bound location proofs.
In the smart tourism scenario, visits to museums and attractions tend to take at least 30 minutes, making this approach viable.
From our initial assumptions on the location types, all the locations have results according to their criteria except for \emph{Comércio}, which we expected would have sufficient diversity of networks.
Thus we reason that 7 minutes is small enough time interval for a viable tourism location proof system.
If we require more time granularity on the creation of time-bound location proofs, some measures can be taken, for example add infrastructure that generates either noise to the network spectrum, \emph{e.g.} a beacon, or deploy a custom AP that dynamically changes its address.

With the positive results of this assessment, that show the feability of the approach, we set out to formalize a model for the data and its operations.

\section{Formal Model}
\label{sec:model}

We now have Wi-Fi traces being collected, from multiple devices, at different locations.
We also did a preliminary assessment.
Now, we to make sense of all this data in a more formal way, it needs to be organized, according to the time of collection, and prepared for use in well-defined location and time determination and verification operations.

\subsection{Data Organization}

We use Relational Algebra~\cite{ElmasriNavatheSham2016} to represent the hotspot data model.
A set of relations\footnotemark~has been defined that represent the system entities and can store information about the collected signals made by the users at different locations.
The model is a formal way to define and verify these relations with respect to their use in computing and evaluating the location evidences.
Table~\ref{table:relations} lists the relations and their attributes and descriptions.
Each relation has a set of attributes that describe the interesting properties for the operations.

\footnotetext{
Relations can also be seen as tables of data, with rows and columns.
This is the terminology usually adopted in relational databases.
However, the model we present abstracts from specific technologies.
}

\begin{table}
\caption{Relations in the model.}
\centering
\label{table:relations}
\begin{tabular}{|c|c|c|}
\hline
\multicolumn{1}{|l|}{\textbf{Relation}} & \textbf{Attributes}   & \textbf{Description} \\ \hline
\multicolumn{1}{|l|}{Obs} &\makecell{id, obsTime, location \\ device, signalType \\ transmitterID} & \makecell{Collected signal} \\ \hline
\multicolumn{1}{|l|}{Locations} & \makecell{id, name, coordinates}& \makecell{Locations or \\ Points of Interest} \\ \hline 
\multicolumn{1}{|l|}{Devices}       & \makecell{id, name, userID}& \makecell{Users mobile \\ devices } \\ \hline
\multicolumn{1}{|l|}{Users}       & \makecell{id, name }& \makecell{Users of the system} \\ \hline
\end{tabular}
\end{table}

\subsection{Time Intervals}
\label{sec:time-windows}

An explicit definition of the temporal property of the model is essential.
We define a precise time-framing that accurately defines boundaries and limits scope and amount of data needed for each operation on the relations.
The proper time-framing to use in the model mainly depends on the application and relevant to the implemented use case and its value is given as a configuration parameter in the system setup.
There are three time intervals in the model.
\begin{itemize}
    \item \emph{Epoch}: The longest time frame which defines the time interval that selects data for the computation of the stable networks at each location or point of interest.
At the start, the system computes the stability of the Wi-Fi networks at each location considering only observations collected within the defined \emph{epoch} time window of the system.
For example, a time interval of 1-week \emph{epoch} means that the system should consider data (observations) collected only during last week to identify the stable Wi-Fi networks at each location.
    \item \emph{Period}: A period is a subdivision of an \emph{epoch} that defines the deadline for the collection of device observations.
For example, a time interval of 1-day \emph{period} defines that the system needs to wait until end of the day to collect observations and then be able to verify the locations of the users.
It means that we consider only data submitted until the end of the \emph{period}, as these data will be most relevant for computing time-bound location proofs.
    \item \emph{Span}: This time interval is defined to represent the accuracy of the produced time-bound location proof.
Upon receiving a location claim from the prover, the system computes the smallest span around the time of the claim ( $t_p$) with additional parameter $\delta$, i.e., the interval is between $t_p-\delta$ and $t_p+ \delta$.
The value of $\delta$ needs to be smaller or equal to the period, but, the ideal is to have the smallest delta possible, so that the location proof can better support the time and location claim made by the prover.
For example, the prover may claim that a device d was at location \textit{Jer\'onimos} at $10:30$'' and the system may only be able to verify that there is evidence that device d was at location \textit{Jer\'onimos} \textit{between} $10:00$ and $11:00$'' or ``$10:30 \pm :30$''.
In this case the $delta$ is 30 minutes.
If more fine grained evidence was available, the claim could be more bound.
For $\delta$ of 10 minutes, the verification could state ``$10:30 \pm :10$''.
\end{itemize}

In summary, the \emph{epoch} is the interval for computing stable networks that provide location, the \emph{period} is the interval for collecting observations, and the span is the smallest interval where evidence was found to verify a location and time claim.
The specific interval sizes -- 1-week \emph{epoch}, 1-day \emph{period}, 30-minute span -- are just illustrative and should be adjusted for the time granularity required by a specific application domain.
However, the following invariant must hold:
$epoch > period > span$.

\subsection{Operations}

The main operations that need to be supported are determining the location and time interval (as small as possible) of a visit.
A visit here is the act of the tourist going to a place to enjoy it.

To support the main operations, we need some auxiliary operations that need to be done with relational algebra.
Table~\ref{table:relationsAlgebraNotations} shows the meaning of the notations used in the algorithms.

\begin{table}
\caption{Relational Algebra notations}
\centering
\label{table:relationsAlgebraNotations}
\begin{tabular}{|c|c|}
\hline
\multicolumn{1}{|l|}{\textbf{Operation}} & \textbf{Description} \\ \hline
\multicolumn{1}{|l|}{$\sigma_{<condition>}(R)$} & \makecell{select tuples from \\ relation R } \\ \hline
\multicolumn{1}{|l|}{$\Pi_{<attribute list>}(R)$} & \makecell{project subset of columns \\ from relation R with no duplicate tuples} \\ \hline
\multicolumn{1}{|l|}{$\tau_{<attribute list>}(R)$} & \makecell{return ordered list of tuples \\ from relation R} \\ \hline

\multicolumn{1}{|l|}{$\leftarrow$} & \makecell{assign \\ operation} \\ \hline
\end{tabular}
\end{table}

\subsection{Location of Visit}

To uniquely estimate the location of a device during a visit, we need to know, beforehand, the identifications of the stable Wi-Fi networks at all locations or points of interests.
Then we use this knowledge to estimate the locations of the users.

\subsubsection{Computing Stable Networks}

This step is done by the system operator before the system is live and is used to identify stable and longer available Wi-Fi network APs at each location within the pre-defined \emph{epoch} time interval of the system.
For better accuracy, the system operator uses multiple devices to gather the Wi-Fi traces.

Algorithm~\ref{alg:computeStableIDs} illustrates, in relational algebra, how to compute the stable network identifications from observations collected within a previous \emph{epoch} time interval.

\algnewcommand\algorithmicforeach{\textbf{for each}}
\algdef{S}[FOR]{ForEach}[1]{\algorithmicforeach\ #1\ \algorithmicdo}

\algnewcommand\algorithmicassert{\textbf{assertTrue}}
\algnewcommand\Assert[1]{\State \algorithmicassert(#1)}%

\algnewcommand\algorithmicassertf{\textbf{assertFalse}}
\algnewcommand\AssertFalse[1]{\State \algorithmicassertf(#1)}%

\begin{algorithm}[tbh]
    \caption{computeStableIDs}
    \label{alg:computeStableIDs}\footnotesize
    \begin{algorithmic}[1] 
        \STATEx{\textbf{Input} $obs$ // set of collected observations}
        \STATEx{\textbf{Input} $e$} // epoch time window
        \STATEx{\textbf{Ouput} $stableIDs$ //stable ids in all locations}
        \STATE{$epochObs \leftarrow \tau_{loc} (\sigma_{obsTime\geq e.start \bigwedge obsTime \leq e.end}(Obs))$}
        \ForEach {$ loc \in \mathcal Locations $}
            \STATE{$locObs \leftarrow \sigma_{location=loc}(epochObs)$}
            \STATE{$devObs \leftarrow \phi $} // start with empty device observations
            \ForEach{$d \in \mathcal Devices$}
                \STATE{$devObs(d) \leftarrow \pi_{transmitterID}(\sigma_{device=d}(locObs))$}  
            \ENDFOR
            \IF{$(NOT devObs.isEmpty())$}
                \STATE{$stableIDs(loc) \leftarrow set.intersection(devObs)$}
            \ENDIF
        \ENDFOR
        \AssertFalse{stableIDs.isEmpty()}
        \Assert{set.intersection(stableIDs).isEmpty()}
        \STATE \textbf{return} $stableIDs$ 
\end{algorithmic}
\end{algorithm}

Observations within the \emph{epoch} time window are selected and ordered by location.
The algorithm takes the observations in each location individually and iterates over each device's observations to collect the unique set of networks IDs.
Then the algorithm computes the intersection between the devices observations to identify the stable networks IDs of the location -- stableIDs(loc).
The algorithm repeats these steps for each location observations to compute the stableIDs of all locations.

\subsubsection{Determining the Location}

When the prover submits a location claim/proof request (as in Figure~\ref{fig:scavenging_timebound_proof}), the system compares the prover submission with the $stableIDs$ to determine the location of the prover.

Algorithm~\ref{alg:locEstimation} illustrates the steps of the location estimation.

\begin{algorithm}[tbh]
  \caption{Prover Location Estimation}
  \label{alg:locEstimation}\footnotesize
  \begin{algorithmic}[1] 
    \STATEx{\textbf{Input} $Obs(p)$ // prover observations}
    \STATEx{\textbf{Input} $stableIDs$ // stable networks at all locations}
    \STATEx{\textbf{Input} $threshold$ // minimum number of detected APs}
    \STATEx{\textbf{Output} $loc$  // estimated prover location}
    \WHILE{$stableIDs$.hasNext()}
        \STATE{$locStableIDs\gets stableIDs.next()$}
        \IF{$(locStableIDs.compare(Obs(p)) \geq threshold)$}
            \STATE{$loc \gets locStableIDs.loc$} \label{alg:wl_4}
            \STATE \textbf{return} $loc$
        \ENDIF
    \ENDWHILE
    \STATE \textbf{return} $loc$ \label{alg:wl_9} // NULL or default location
\end{algorithmic}
\end{algorithm}

The algorithm starts by iterating over all stable networks present in the $stableIDs$ to find the \emph{StableIDs} that best-matched the prover's submitted observations.
This location will be the estimated location of the prover device.

\subsection{Time of Visit}

After estimating the location of the prover, the time of visitation at a location can be determined.
This requires sets of observations, \textit{i.e.}, Wi-Fi traces, reported by other users (\emph{witnesses}) that happen to be available at the prover's location during the same time span.
This is done in the model by computing \emph{volatileIDs}, containing a set of network APs resulted from the intersection of the observations reported by the witnesses, excluding those that appeared in the \emph{stableIDs} of the location.
Networks that are stable over long periods of time do not contribute to the location's entropy and, therefore, are not suitable to determine time of visitation.

\subsubsection{Computing Volatile Networks}

Algorithm~\ref{alg:computeVolatileIDs} illustrates how to compute the \emph{volatileIDs}, considering observations from the witness users.

\begin{figure}
	\centering
	\includegraphics[width=0.55\columnwidth]{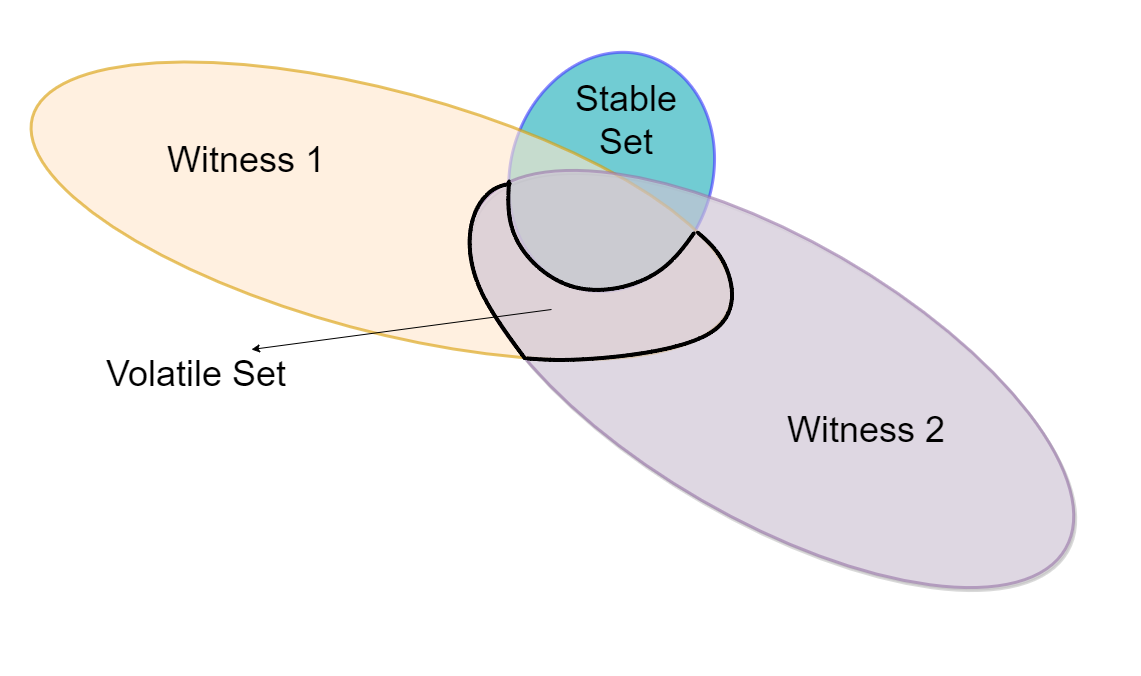}
	\caption{Computing the \emph{VolatileIDs} from two witness observation sets.}
	\label{fig:venn_volatile_set}
\end{figure}

\begin{algorithm}[tbh]
    \caption{computeVolatileIDs}
    \label{alg:computeVolatileIDs}\footnotesize
    \begin{algorithmic}[1] 
        \STATEx{\textbf{Input} $obs$ // set of collected observations}
        \STATEx{\textbf{Input} $c$} // location claim
        \STATEx{\textbf{Input} $s$} // span time window
        \STATEx{\textbf{Input} $threshold$} // minimum number of witnesses
        \STATEx{\textbf{Ouput} $volatileIDs$}
         \STATE{$spanObs \leftarrow \sigma_{(obsTime\geq s.start \bigwedge obsTime \leq s.end) \bigwedge (loc=c.loc)}(Obs)$}
         \STATE{$witnesses \leftarrow \pi_{device}(spanObs)$}
         \ForEach{$w \in \mathcal Witnesses $}
            \STATE{$witnessObs(w) \leftarrow \pi_{transmitterID}(\sigma_{device=w}(spanObs))$}
         \ENDFOR
         \Assert{witnessObs.size()$\geq$ threshold}
        \STATE{$volatileIDs \leftarrow set.intersection(witnessObs) - stableIDs$}
        \STATE \textbf{return} $volatileIDs$ 
\end{algorithmic}
\end{algorithm}
Figure~\ref{fig:venn_volatile_set} presents a Venn diagram that illustrates the computation of the \emph{volatileIDs}.

\subsubsection{Determining the Time Interval}

We hypothesize that dividing the time span of the prover's location claim into smaller intervals and verify each of those intervals individually can help to pinpoint, with more certainty, the time interval at which the prover was at the claimed location.
Algorithm~\ref{alg:timeBoundProof} illustrates the computation with list of possible time spans we call them \emph{deltas}.
\begin{algorithm}[tbh]
    \caption{timeBoundProof}
    \label{alg:timeBoundProof}\footnotesize
    \begin{algorithmic}[1] 
        \STATEx{\textbf{Input} $obs$ // set of collected observations}
        \STATEx{\textbf{Input} $claim$} // location claim
        \STATEx{\textbf{Input} $Deltas[120~min, 60~min, 30~min, 15~min, 10~min, 5~min, 1~min, 0~min]$}
        \STATEx{\textbf{Ouput} $proof (TRUE~/~FALSE)$}
        \STATEx{\textbf{Ouput} $proofDelta $}
        \STATE{$proof \leftarrow FALSE$}
        \STATE{$proofDelta \leftarrow 0 $}
        \ForEach{$delta \in \mathcal Deltas $} // from larger delta to smaller delta
            \STATE{$span \leftarrow (claim.time-delta, claim.time+delta)$}
            \STATE{$volatileIDs \leftarrow computeVolatileIDs(obs, claim, span)$}
            \STATE{$proofSet \leftarrow set.intersection (volatileIDs , claim.evidence) $}
            \IF{$(proofSet = \phi)$} // intersection was empty
                \STATE{$break$}
            \ELSE
                \STATE{$proofDelta = delta$}
            \ENDIF
         \ENDFOR
         
        \IF{$(proofDelta > 0)$}
            \STATE{$proof \leftarrow TRUE$}
        \ENDIF
        
        \STATE \textbf{return} $proof, proofDelta$ 
\end{algorithmic}
\end{algorithm}
Given the list of \emph{deltas}, the algorithm starts by computing the span time for each delta in the list.
As mentioned, a span is the time window around the time of the location claim ($t_p$) requested by the prover with additional delta that makes up the interval between $t_p-\delta$ and $t_p+ \delta$.
The algorithm then computes the volatileIDs with respect to the selected \emph{delta}.
This step is performed by calling Algorithm~\ref{alg:computeVolatileIDs}.
Then the intersection between the volatileIDs and the location evidence in the prover claim is computed.
A non-empty set result from the intersection indicates that the system can produce proof of location for this time span (\emph{proofDelta}).
Then the algorithm iterates over all \emph{deltas} to find the smallest that can be used for producing the location proof.
The result is TRUE for the proof and the smallest time span found for the location proof.
In case all the \emph{deltas} gave empty results, then the algorithm returns FALSE proof, indicating that the location proof cannot be produced.

\subsection{Summary}

We presented the model for the formal definition of the data relations and the algorithms that use them to compute the relevant network sets, and to perform the operations to determine the location and the smallest time interval where the presence verification is possible.

\section{Conclusion}
\label{sec:conclusion}

The premise of this work was that there is a large number of publicly available Wi-Fi hotspots in a city, and that some of these are long-lived and others are short-lived.
We investigated how the \textit{hotspot observations} can be combined to \textit{detect the location} and to \textit{prove the time when the location was visited} and showed how the intersection of observation sets by other users -- witnesses -- can corroborate the location claims and produce credible location proofs.

The results of the \textit{field experiment} made in 6 locations of Lisbon over a period spanning 6 months, was collected as a dataset called \textit{LXspots}.
The data was assessed in a smart tourism context and we have shown that the approach is viable and worth implementing in practice.
The assessment also lay the groundwork that allowed the development of the formal data model and algorithms for determining the location and time interval of a tourist visit.
The results show the feasibility of a Wi-Fi scavenger approach.
The developed model can be extended to include other kinds of volatile network signals, such as nearby Bluetooth devices, to further improve the produced time-bound location proofs.


\section*{Acknowledgements}
\label{sec:ack}

This work was supported by national funds through Funda\c{c}\~{a}o para a Ci\^{e}ncia e a Tecnologia (FCT) with reference UIDB/50021/2020 (INESC-ID) and through project with reference PTDC/CCI-COM/31440/2017 (SureThing).

\clearpage

\bibliographystyle{unsrt}  
\bibliography{paper}

\end{document}